\documentclass[aps,prl,twocolumn,showpacs]{revtex4}
\usepackage[dvips]{graphicx}
\usepackage{amsmath}
\usepackage{times}

\begin{document}
\author{ %
Yan V. Fyodorov$^1$ and Dmitry V. Savin$^2$}
\affiliation{ 
$^1$Department of Mathematical Sciences, Brunel University,
Uxbridge UB83PH, United Kingdom \\ %
$^2$Fachbereich Physik, Universit\"at Duisburg-Essen, 45117 Essen,
Germany}

\title{Statistics of Impedance, Local Density of States, and
Reflection in Quantum Chaotic Systems with Absorption.}
\date{September 6, 2004}


\begin{abstract}
We are interested in finding the joint distribution function of
the real and imaginary parts of the local Green function for a system
with chaotic internal wave scattering and a uniform energy loss
(absorption). For a microwave cavity attached to a single-mode
antenna the same quantity has a meaning of the complex cavity
impedance. Using the random matrix approach, we relate its
statistics to that of the reflection coefficient and scattering
phase and provide exact distributions for systems with $\beta{=}2$
and $\beta{=}4$ symmetry class. In the case of $\beta{=}1$ we
provide an interpolation formula which incorporates all known
limiting cases and fits excellently available experimental data as
well as diverse numeric tests.
\end{abstract}


\maketitle  

Characterising statistical fluctuations of physical observables in
quantum systems with underlying chaotic classical dynamics remains
a very active field of research in theoretical and experimental
physics. A considerable progress in understanding the phenomenon
was underpinned by revealing the apparent universality of the
fluctuations in systems of very diverse microscopic nature,
ranging from atomic nuclei and Rydberg atoms in strong external
fields, to complex molecules, quantum dots, and mesoscopic
samples, see e.g. \cite{Stoeckmann}. From theoretical side, the
universality allows one to exploit the random matrix theory (RMT)
as a powerful tool for analysis of generic features of the energy
spectra of such systems \cite{Guhr1998,Alhassid2000}.

In many atomic, molecular, and mesoscopic systems the quantity
which is readily obtained experimentally is the absorption
spectrum for transitions from a given initial state $|g\rangle$ to
highly excited chaotic states at the energy $E$. For
high-resolution experiments chaotic spectra consists of
well-resolved narrow resonance peaks and one can, in principle,
study statistics of the peak heights and widths as well as that of
spacings between consecutive peaks. Most frequently, however, the
absorption spectra look practically continuous due to both
inevitable level broadening and finite experimental resolution.
Then the relevant statistics is the distribution and correlation
functions of the absorption probability $\sigma(E)$ (also known as
the strength function of the dipole operator $\hat{\mu}$) which in
the simplest situation of uniform level broadening $\Gamma$ is
given by $\sigma(E)\propto\textrm{Im}\langle
g|\hat{\mu}\,\hat{G}(E)\,\hat{\mu}|g\rangle$, see e.g.
Ref.~\cite{Fyodorov1998i} and discussion therein.
Assuming the  validity of the RMT, the problem then amounts to
studying statistical properties of the resolvent (Green function)
operator $\hat{G}(E)\equiv(E+i\Gamma/2-\hat{H})^{-1}$ associated
with the random matrix $\hat{H}$ which replaces the actual chaotic
Hamiltonian. In particular, the imaginary part of the diagonal
entries of $\hat{G}(E)$ is well known in the solid state physics
as the \textit{local density of states} (LDoS) and in this
capacity its statistics enjoyed many studies
\cite{Fyodorov1998i,Efetov1993,Beenakker1994,Mirlin1994,Taniguchi1996}.

From the experimental point of view the same universality, which
makes the use of the RMT legitimate, provides one with an
attractive possibility to employ simple model systems for
analyzing generic statistics of the fluctuating quantities. One of
such systems which proved to be an ideal playground for
investigating a variety of quantum chaos phenomena are various
microwave billiards \cite{Stoeckmann}. The billiards are realized
as resonators in a form of electromagnetic cavities fed up with
the radiation of the wavelength small compared with the
characteristic size of the enclosure and shaped to ensure the
chaoticity of internal scattering. The cavities are coupled to
transmission lines or to waveguides which are used to inject
electromagnetic waves into the system as well as to collect the
outgoing waves. The adequate description of the experiment is then
achieved in terms of the scattering matrix $S$ relating amplitudes
of incoming and outgoing waves.

High-resolution experiments are usually performed in
low-temperature (superconducting) cavities with very high quality
factor \cite{Dembowski2002}. Majority of the experiments are,
however, done at room temperatures
\cite{Mendez-Sanchez2003,Hemmady2004,Barthelemy2004,Kuhl2004}.
Inevitable energy losses (absorption) leading to uniform
broadening of the resonances play, therefore, an important role
and have to be taken into account adequately when describing the
experiments theoretically.

In the present paper we are going to concentrate on the simplest
case of a single one-channel antenna experiment. To be able to
employ the RMT methods, it is conventional to represent the
resonant part of the scattering matrix $S$ in the following form
(see e.g. Ref.~\cite{Fyodorov1997}):
\begin{equation}\label{S}
S(E) = \frac{1-iK(E)}{1+iK(E)}\equiv\sqrt{r}e^{i\theta} \,,
\end{equation}
where $K(E)=V^{\dagger}\hat{G}(E)V$ is the so-called $K$ matrix.
The Hamiltonian $\hat{H}$ of the \textit{closed} chaotic cavity
gives rise to $N$ eigenfrequencies $E_n$ characterized in the
relevant range of the scattering energy $E$ by the mean level
spacing $\Delta$. The column-vector $V$ describes the
energy-independent amplitudes coupling the corresponding
eigenmodes to the propagating mode in the antenna. We see again
that the study of statistical properties of the scattering matrix
amounts to knowing statistics of the diagonal element $G_{11}$ of
the Green function of the closed system in some basis. In fact, in
the present context the function $iK\equiv Z$ has the direct
physical meaning of the electric impedance $Z$ of the cavity which
relates linearly voltages and currents at the antenna port
\cite{Hemmady2004}.

Without absorption $\Gamma=0$ and the scattering matrix is
unimodular: $r\equiv 1$. At finite absorption, $\Gamma\!>\!0$, the
reflection coefficient $r$ and the scattering phase $\theta$ have
nontrivial distributions, which have been recently measured in
experiment \cite{Mendez-Sanchez2003,Kuhl2004}. On the other hand,
universal fluctuations of both real and imaginary parts of the
cavity impedance $Z$ have been recently investigated
experimentally in \cite{Hemmady2004}. Since the impedance matrix
$Z$ is related to eigenmodes and eigenfrequencies of the closed
cavity, the study of $Z$ is in some sense complementary to that of
$S$.

Various statistics related to the scattering matrix of chaotic
systems with losses were subject of a number of recent papers
\cite{Ramakrishna2000,Kogan2000,Beenakker2001,Savin2003i,Fyodorov2003i,Fyodorov2004}.
Explicit analytical results were available, however, only for the
simplest case of systems with no time-reversal invariance (TRI)
corresponding to the so-called $\beta{=}2$ symmetry class of the
RMT. At the same time, majority of the billiard-type experiments
is performed in systems which are time-reversal invariant
($\beta{=}1$ symmetry class of the RMT). Similar situation holds
for the statistics of the local Green function, in particular,
LDoS. For the $\beta{=}2$ case the corresponding expressions were
obtained by various methods in
\cite{Efetov1993,Beenakker1994,Mirlin1994,Andreev1995}. An attempt
\cite{Taniguchi1996} to provide an expression for the LDoS
distribution for the $\beta{=}1$ case can not be considered as
particularly successful, as the general expression was given in a
form of an intractable fivefold integral. Finally, it is worth
mentioning the existence of the $\beta{=}4$ symmetry class
describing time-reversal invariant chaotic systems with
half-integer spin. This situation may occur in quantum dots with
strong spin-orbit scattering \cite{Aleiner2001}, in Rydberg atoms
driven by microwave fields \cite{Sacha2001}, and can be
efficiently simulated in some other models of quantum chaotic
systems \cite{Bolte2003}.

The fundamental quantity which determines the full statistics of
$S$ or $Z$ is the joint distribution function ${\mathcal{P}}(u,v)$
of the real $u=\textrm{Re\,}K$ and imaginary $v=-\textrm{Im\,}K>0$
parts of $K$. Generally, one can always write
$K=\kappa(N\Delta/\pi)G_{11}$ in the RMT. The effective coupling
constant $\kappa=\pi\|V\|^2/N\Delta>0$ enters the $S$ matrix
statistics only through the so-called transmission coefficient
$T\equiv1-|\overline{S}|^2$ ($=4\kappa/(1+\kappa)^2$ in the middle
of the spectrum, $E=0$), see e.g. Ref.~\cite{Fyodorov1997} for
details.

A convenient starting point of our analysis is the observation
that the distribution ${\mathcal{P}}(u,v)$ must always have the
following general form:
\begin{equation}\label{P(u,v)}
{\mathcal{P}}(u,v) = \frac{1}{2\pi v^2}P_0(x) \,,
\end{equation}
with $x=(u^2+v^2+1)/2v>1$. It initially emerged in
\cite{Mirlin1994i} in the course of explicit calculations for the
$\beta{=}2$ symmetry class, but neither origin nor generality of
such a form were appreciated. Here we show that Eq.~(\ref{P(u,v)})
is the direct consequence of two fundamental properties of the
so-called ``perfect coupling'' case $T{=}1$: (i) the statistical
independence of the $S$ matrix modulus $r\equiv\frac{x-1}{x+1}$
and its phase $\theta$; and (ii) the uniform distribution of
$\theta\in(0,2\pi)$. Both these properties can be verified using
the methods of Ref.~\cite{Brouwer1997ii}. The joint distribution
$P(x,\theta)$ then factorizes to $P_0(x)/2\pi$. Choose now new
variable
$y\equiv\frac{\textrm{Re\,}S}{\textrm{Im\,}S}=\cot(\theta)$
instead of $\theta$, so that $|d\theta/dy|=(1+y^2)^{-1}$. Noticing
that $y=(u^2+v^2-1)/2u$ and evaluating the corresponding Jacobian
$|\partial(x,y)/\partial(u,v)|=(1+y^2)/v^2$ of the transformation
from $x$ and $y$ to $u$ and $v$, we come after a simple
calculation to (\ref{P(u,v)}).

The explicit form of $P_0(x)$ at arbitrary absorption for various
symmetry classes will be given and discussed below. Having
$P_0(x)$ at our disposal, it is immediate to find the distribution
of the imaginary part $v$ (the LDoS normalized for
convenience to have the unit mean value):
\begin{equation}\label{P(v)}
{\mathcal P}_{v}(v) = \frac{\sqrt{2}}{\pi
v^{3/2}}\int_0^{\infty}\!\!dq\,
P_0\left[q^2+\frac{1}{2}\Bigl(v+\frac{1}{v}\Bigr)\right]\,,
\end{equation}
The distribution is normalized to 1 and has the first moment
unity,
$\langle{v}\rangle\equiv\int_{0}^{\infty}\!\!dv\,v{\mathcal P}(v)=1$,
automatically due to invariance of the integrand with respect to
the change $v\to1/v$. Similarly, one can find the distribution of
the real part $u$ to be:
\begin{equation}\label{P(u)}
{\mathcal P}_{u}(u) =
\frac{1}{2\pi\sqrt{u^2+1}}\int_0^{\infty}\!\!dq\,
P_0\left[\frac{\sqrt{u^2+1}}{2}\Bigl(q+\frac{1}{q}\Bigr)\right]\,.
\end{equation}
Although $u$ has no direct physical meaning in the context of
solid state mesoscopic systems, both ${\mathcal P}_{u}(u)$ and
${\mathcal P}_{v}(v)$ are directly measurable in microwave cavities
\cite{Hemmady2004}.

Let us now discuss the explicit forms of $P_0(x)$ for various
symmetry classes: $\beta{=}1,2,4$. For the simplest case of broken
TRI ($\beta{=}2$) an exact result is available at arbitrary values
of the dimensionless absorption strength
$\gamma\equiv2\pi\Gamma/\Delta$
\cite{Fyodorov2003i,Savin2003i,Beenakker2001}. Scaling the
absorption parameter for the subsequent use as
$\alpha\equiv\gamma\beta/2$, we can represent the $\beta{=}2$
result as follows:
\begin{equation}\label{P(x)}
P_0(x) = \frac{{\mathcal N}_{\beta}}{2}
\left[A\,\left(\alpha(x+1)/2\right)^{\beta/2} + B\right]
e^{-\alpha(x+1)/2} \,,
\end{equation}
with $\alpha$-dependent constants $A\equiv e^{\alpha}-1$ and
$B\equiv1+\alpha-e^{\alpha}$ and the normalization constant
${\mathcal N}_2=1$.

For the case $\beta{=}4$ the exact form became available very
recently \cite{unpub} by exploiting important advances in the RMT
\cite{BorodinStrahov}. The explicit derivation will be given
elsewhere, the final result being \cite{unpub}:
\begin{equation}\label{P(x)gse}
P_0^{\textrm{gse}}(x) = \widetilde{P}_0^{\textrm{gue}}(x) +
C(x,\gamma) e^{-\gamma x}
\int_{0}^{\gamma}\!\!dt\frac{\sinh{t}}{t} \,,
\end{equation}
where $\widetilde{P}_0^{\textrm{gue}}(x)$ is the distribution
(\ref{P(x)}) for $\beta{=}2$ taken, however, at
$\alpha\!=\!2\gamma$ and
$C(x,\gamma)\equiv\gamma^2(x+1)^2/2-\gamma(\gamma+1)(x+1)+\gamma$.

Unfortunately, for the most interesting case $\beta{=}1$ the
explicit formula for $P_0(x)$ is not available yet, apart from the
limiting cases of weak or strong absorption:
\begin{equation}\label{P(x)lim}
P_0(x) \simeq \left\{ \begin{array}{ll}
\frac{\alpha^{\beta/2+1}}{2\Gamma(\beta/2+1)}
\left(\frac{x+1}{2}\right)^{\beta/2}e^{-\alpha(x+1)/2}\,, &
\gamma\ll1 \\[2ex]
\alpha\,e^{-\alpha(x-1)/2}\,, & \gamma\gg 1.
\end{array}\right.
\end{equation}%
The first line here results from the relation
$\frac{2}{x+1}\!=\!1-r\!\approx\!\gamma\tau$ at $\gamma\ll1$
between the reflection coefficient and the (dimensionless)
time-delay $\tau$, see \cite{Ramakrishna2000,Beenakker2001}. The
time-delay distributions are known \cite{Fyodorov1997,Gopar1996}
for all $\beta{=}1,2,4$:
${\mathcal P}_{\tau}(\tau)=[(\beta/2)^{\beta/2}/\Gamma(\beta/2)]
\tau^{-\beta/2-2}e^{-\beta/2\tau}$. In the opposite case
$\gamma\gg 1$, the known limiting Rayleigh distribution
\cite{Kogan2000} $P(r)\simeq(\gamma\beta/2)e^{-r\gamma\beta/2}$
yields the second line in Eq.(\ref{P(x)lim})

In the absence of the general expression for $\beta{=}1$ a natural
idea is  to try to invent a formula interpolating between the
known limiting cases \cite{Kuhl2004}. We suggest Eq.~(\ref{P(x)})
to be the appropriate natural candidate, with the normalization
constant being
${\mathcal N}_{\beta}=\alpha/(A\Gamma(\beta/2{+}1,\alpha)+Be^{-\alpha})$,
where
$\Gamma(\nu,\alpha)=\int_{\alpha}^{\infty}\!\!dt\,t^{\nu-1}e^{-t}$.
Indeed, such a form reproduces correctly Eq.~(\ref{P(x)lim}) as
both limits are determined solely by the first (universal) term in
(\ref{P(x)}). One needs, however, to keep $B$ in order to handle
properly the case of moderate absorption ($\alpha\sim1$).

\begin{figure}[t]
\includegraphics[width=0.45\textwidth]{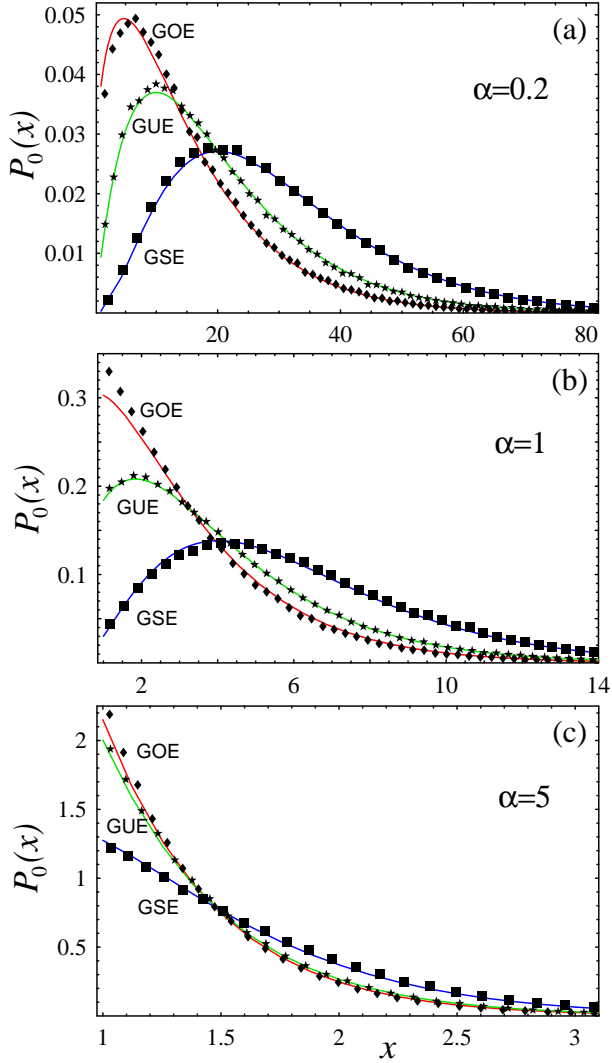}
\caption{The distribution $P_0(x)$, Eq.~(\ref{P(x)}), at different
values of the absorption parameter $\alpha{=}\beta\pi\Gamma/\Delta$.
Solid lines correspond to the $\beta{=}1$\,(GOE) and 2\,(GUE)
cases and to the exact $\beta{=}4$\,(GSE) result,
Eq.~(\ref{P(x)gse}). Symbols stand for the numerics done for 400
realizations of 500$\times$500 random matrices.}
\end{figure}

Figure 1 shows results of numerical simulations with random
matrices drawn from the Gaussian orthogonal (GOE, $\beta{=}1$),
unitary (GUE, $\beta{=}2$) or symplectic (GSE, $\beta{=}4$)
ensemble. The overall agreement of Eq.~(\ref{P(x)}) with the data
at $\beta{=}1$ is nearly as good as for the exact cases
$\beta{=}2,4$. Another check has been performed in
Ref.~\cite{Kuhl2004} which measured distributions of the
reflection coefficient and the scattering phase in the broad range
of system parameters and found very good agreement with the
corresponding expressions following from Eq.~(\ref{P(x)}).

We discuss now the behavior of ${\mathcal P}_{v}(v)$ and
${\mathcal P}_{u}(u)$. Performing the integration in
Eq.~(\ref{P(v)}), we arrive for $\beta{=}2$ at
\begin{eqnarray}\label{P(v)gue}
{\mathcal P}_{v}^{\textrm{gue}}(v) &=& (\gamma/16\pi)^{1/2}v^{-3/2}
\exp[-\gamma(v+v^{-1})/4] \nonumber  \\
&&\times [ 2\cosh\frac{\gamma}{2} +
(v+v^{-1}-2/\gamma)\sinh\frac{\gamma}{2} ],
\end{eqnarray}
which is exactly the LDoS distribution obtained earlier in
\cite{Efetov1993}. For the case $\beta{=}1$, we get the following
result:
\begin{equation}\label{P(v)goe}
{\mathcal P}_{v}^{\textrm{goe}}(v) =
\frac{{\mathcal N}_1\,e^{-a}}{\pi\sqrt{2\gamma}\,v^{3/2}}
\bigl(A[K_0(a)+K_1(a)]a + \sqrt{\pi}Be^{-a}\bigr)\,,
\end{equation}
with $a\equiv\frac{\gamma}{16}(\sqrt{v}+1/\sqrt{v})^2$ and
$K_{\nu}(z)$ being the MacDonald function. It is instructive to
consider the asymptotic behavior of these functions in the limits
of small or large absorption. At $\gamma\ll1$, the distribution
${\mathcal P}_{v}(v)$ becomes very broad having the maximum at
$v\sim\gamma$, a power-law bulk behavior and an exponential cutoff
at the far tails:
\begin{equation}\label{P(v)weak}
{\mathcal P}_{v}(v) \propto \left\{
\begin{array}{ll}
\alpha^{(1+\beta)/2}v^{-(3+\beta)/2}e^{-\alpha/4v}\,,\quad &
\!v\ll\alpha\\[1ex]
\alpha^{1/2}v^{-3/2}\,, \hfill\alpha\ll  &\!v \ll 1/\alpha\\[1ex]
\alpha^{(1+\beta)/2}v^{-(3-\beta)/2}e^{-\alpha v/4}\,,\quad &
\!\!\!1/\alpha\ll v \,,
\end{array}\right.
\end{equation}
where constants $\sim1$ are omitted. This result can be physically
interpreted in the single-level approximation
\cite{Efetov1993,Taniguchi1996}, when the bulk and tail behavior
is governed by spectral and wave function fluctuations,
respectively. As $\gamma$ increases, a number of levels
contributing to $v$ grows as $\sim\gamma$, so that ${\mathcal
P}_{v}(v)$ tends to the limiting Gaussian distribution
\begin{equation}\label{P(v)strong}
{\mathcal P}_{v}(v) = \sqrt{\frac{\alpha}{4\pi v^{3}}} \exp\left[
-\frac{\alpha}{4}\Bigl(\sqrt{v}-\frac{1}{\sqrt{v}}\Bigr)^2\right]
\,,
\end{equation}
which has a peak at $v\sim1$ of the width
$\propto1/\sqrt{\gamma}\ll1$, in agreement with the earlier result
\cite{Taniguchi1996}.

As to the distribution ${\mathcal P}_u(u)$, equation
(\ref{P(u)}) leads after the integration to the following exact
$\beta=2$ result:
\begin{equation}\label{P(u)gue}
{\mathcal P}_{u}^{\textrm{gue}}(u) = \frac{\gamma}{2\pi} \left[
\sinh\frac{\gamma}{2}K_0\Bigl(\frac{\gamma\tilde{u}}{2}\Bigr) +
\frac{\cosh\frac{\gamma}{2}}{\tilde{u}}
K_1\Bigl(\frac{\gamma\tilde{u}}{2}\Bigr) \right]\,,
\end{equation}
where $\tilde{u}\equiv\sqrt{u^2+1}$. Integrating the interpolation
formula for the case $\beta=1$, we obtain
\begin{equation}\label{P(u)goe}
{\mathcal P}_{u}^{\textrm{goe}}(u) = \frac{{\mathcal N}_1
e^{-\gamma/4}}{2\pi\tilde{u}} \left[
\frac{A}{2}\sqrt{\frac{\gamma}{4}}\,
D\Bigl(\frac{\tilde{u}}{2}\Bigr) + B
K_1\Bigl(\frac{\gamma\tilde{u}}{4}\Bigr) \right]\,,
\end{equation}
where
$D(z){\equiv}\int_0^{\infty}\!\!dq\sqrt{1+z(q+q^{-1})}e^{-\gamma
z(q+q^{-1})/4}$ is introduced for convenience. The limiting forms
of ${\mathcal P}_u(u)$ at weak and strong absorption follow
readily as
\begin{equation}\label{P(u)lim}
{\mathcal P}_{u}(u) \simeq \left\{ \begin{array}{ll}
\pi^{-1}(1+u^2)^{-1}\,, & \alpha\ll1 \\[2ex]
\sqrt{\alpha/4\pi}\,e^{-\alpha u^2/4}\,, & \alpha\gg1\,,
\end{array} \right.
\end{equation}
and describe a crossover from the Lorentzian to Gaussian
distribution as absorption grows. This type of behavior as well as
the trend of ${\mathcal P}_v(v)$ to the Gaussian
(\ref{P(v)strong}) was recently observed in the experimental study
of the cavity impedance \cite{Hemmady2004}.

Finally, let us mention that the case of nonperfect coupling,
$T<1$, can be mapped \cite{Brouwer1995,Savin2001} onto that of
perfect one making use of the parametrization \cite{Mello1985}
$S_0=(S-\sqrt{1\!-\!T})/(1-\sqrt{1\!-\!T}S)$. Here $S_0$ is the
scattering matrix of the system  in the perfect coupling case. Now
$x$ and $\theta$ do correlate and, after the evaluation in
parametrization (\ref{S}) of the corresponding Jacobian, the joint
distribution $P(x,\theta)$ is found to be
\begin{equation}\label{P(x,theta)}
P(x,\theta) = \frac{1}{2\pi}
P_0(xg-\sqrt{(x^2-1)(g^2-1)}\cos\theta) \,,
\end{equation}
with $g\equiv2/T-1$. Complementary to Eq.~(\ref{P(u,v)}), equation
(\ref{P(x,theta)}) provides an access to scattering observables.
The integration there over $x$ yields the scattering phase
distribution. In particular, when absorption vanishes,
$x\to\infty$ ($r\to1$) and $P_0(x)\to\delta(1/x)$, giving readily
$P(x,\theta)=\rho(\theta)\delta(1/x)$, with the phase density
$\rho(\theta)=[2\pi(g-\sqrt{g^2-1}\cos\theta)]^{-1}$ found earlier
\cite{Savin2001}. As another example, the distribution of the
reflection coefficient in terms of $P_0(x)$ is given at arbitrary
coupling by (cf. Eq.(5) in \cite{Fyodorov2003i} and see also
\cite{Kuhl2004,Savin2003i} in this respect):
\begin{equation}\label{P(r)}
P_{r}(r) = \int_0^{2\pi}\!\!\!\frac{d\theta\,\pi^{-1}}{(1-r)^2}P_0
\Bigl[\frac{2(g-\sqrt{g^2-1}\sqrt{r}\cos\theta)}{1-r}-g\Bigr] \,.
\end{equation}

In conclusion, although rigorous analytical treatment of the
$\beta{=}1$ case remains a theoretical challenge it is worth
stressing, however, that the suggested interpolation formulas
should be sufficient for the most of practical purposes of
comparison to the experimental/numerical data. Moreover, all
physically interesting limiting cases, as e.g.
Eqs.~(\ref{P(v)weak}), (\ref{P(v)strong}) and (\ref{P(u)lim}), are
already reproduced from the exact limiting statistics
(\ref{P(x)lim}). At the same time, it is important to understand
that an excellent performance of the interpolation formula
(\ref{P(x)}) for $\beta{=}1$ is nothing else as a lucky
coincidence. Indeed, applying the same formula for the $\beta{=}4$
case, we have found that apart from well reproduced limits of weak and
strong absorption, an agreement with numerics at intermediate
values $\gamma\sim 1$ turns out to be by far not as good as in the
GOE case.

We are grateful to S.~Anlage, P.~Brouwer, U.~Kuhl, A.~Mirlin,
H.-J.~Sommers and E.~Strahov for useful communications at various
stages of this work. The financial support by the SFB/TR 12 der
DFG is acknowledged with thanks.


\begin{thebibliography}{34}

\bibitem{Stoeckmann}
H.-J. St{\"{o}}ckmann, \emph{Quantum Chaos: An Introduction}
(Cambridge University Press, Cambridge,1999).

\bibitem{Guhr1998}
T.~Guhr, A.~M\"{u}ller-Groeling, and H.~A. Weidenm{\"{u}}ller,
Phys. Rep. \textbf{299}, 189 (1998).

\bibitem{Alhassid2000}
Y.~Alhassid, Rev. Mod. Phys. \textbf{72}, 895 (2000).

\bibitem{Fyodorov1998i}
Y.~V. Fyodorov and Y.~Alhassid, Phys. Rev. A \textbf{58}, R3375
(1998).

\bibitem{Efetov1993}
K.~B. Efetov and V.~N. Prigodin, Phys. Rev. Lett. \textbf{70},
1315 (1993).

\bibitem{Beenakker1994}
C.~W.~J. Beenakker, Phys. Rev. B \textbf{50}, 15170 (1994).

\bibitem{Mirlin1994}
A.~D. Mirlin and Y.~V. Fyodorov, Europhys. Lett. \textbf{25}, 669
(1994).

\bibitem{Taniguchi1996}
N.~Taniguchi and V.~N. Prigodin, Phys. Rev. B \textbf{54}, 14305
(1996).

\bibitem{Dembowski2002}
C.~Dembowski \emph{et al.}, Phys. Rev. Lett. \textbf{89}, 064101 (2002).

\bibitem{Mendez-Sanchez2003}
R.~A. M\'endez-S\'anchez \emph{et al.},
Phys. Rev. Lett.  \textbf{91}, 174102 (2003).

\bibitem{Hemmady2004}
S.~Hemmady \emph{et al.},
\emph{e-print} cond-mat/0403225;
X.~Zheng \emph{et al.},
\emph{e-prints} cond-mat/0408317 and cond-mat/0408327.

\bibitem{Barthelemy2004}
J.~Barth{\'{e}}lemy, O.~Legrand, and F.~Mortessagne,
\emph{e-prints} cond-mat/0401638 and cond-mat/0402029.

\bibitem{Kuhl2004}
U.~Kuhl \emph{et al.}, \emph{e-print} cond-mat/0407197.

\bibitem{Fyodorov1997}
Y.~V. Fyodorov and H.-J. Sommers, J. Math. Phys. \textbf{38}, 1918
(1997).

\bibitem{Ramakrishna2000}
S.~A. Ramakrishna and N.~Kumar, Phys. Rev. B \textbf{61}, 3163 (2000).

\bibitem{Kogan2000}
E.~Kogan, P.~A. Mello, and H.~Liqun, Phys. Rev. E \textbf{61}, R17
(2000).

\bibitem{Beenakker2001}
C.~W.~J. Beenakker and P.~W. Brouwer, Physica E \textbf{9}, 463
(2001).

\bibitem{Savin2003i}
D.~V. Savin and H.-J. Sommers, Phys. Rev. E \textbf{68}, 036211
(2003).

\bibitem{Fyodorov2003i}
Y.~V. Fyodorov, JETP Lett. \textbf{78}, 250 (2003).

\bibitem{Fyodorov2004}
I. Rozhkov \emph{et al.}, Phys. Rev. E \textbf{68}, 016204 (2003);
Y.~V. Fyodorov and A.~Ossipov, Phys. Rev. Lett. \textbf{92},
084103 (2004).

\bibitem{Andreev1995}
A.~V. Andreev and B.~D. Simons, Phys. Rev. Lett. \textbf{75}, 2304
(1995).

\bibitem{Aleiner2001}
I.~L. Aleiner and V.~I. Fal{'}ko, Phys. Rev. Lett. \textbf{87},
256801 (2001).

\bibitem{Sacha2001}
K.~Sacha and J.~Zakrzewski, Phys. Rev. Lett. \textbf{86}, 2269
(2001).

\bibitem{Bolte2003}
J.~Bolte and J.~Harrison, J. Phys. A \textbf{36}, 2747 (2003).

\bibitem{Mirlin1994i}
A.~D. Mirlin and Y.~V. Fyodorov, Phys. Rev. Lett. \textbf{72}, 526
(1994); J. Phys. I \textbf{4}, 655 (1994).

\bibitem{Brouwer1997ii}
P.~W. Brouwer and C.~W.~J. Beenakker, Phys. Rev. B \textbf{55},
4695 (1997).

\bibitem{Gopar1996}
V.~A. Gopar, P.~A. Mello, and M.~B{\"{u}}ttiker, Phys. Rev. Lett.
\textbf{77}, 3005 (1996).

\bibitem{Brouwer1995}
P.~W. Brouwer, Phys. Rev. B \textbf{51}, 16878 (1995).

\bibitem{Savin2001}
D.~V. Savin, Y.~V. Fyodorov, and H.-J. Sommers, Phys. Rev. E
\textbf{63}, 035202(R) (2001).

\bibitem{Mello1985}
P.~A. Mello, P.~Pereyra, and T.~Seligman, Ann. Phys. \textbf{161},
254 (1985).

\bibitem{unpub} Y.~V. Fyodorov, unpublished.

\bibitem{BorodinStrahov}
A.~Borodin and E.~Strahov, \emph{e-print} math-ph/0407065.

\end{thebibliography}
%

\end{document}